\documentstyle[12pt,psfig]{article}

\newcommand{\drawsquare}[2]{\hbox{%
\rule{#2pt}{#1pt}\hskip-#2pt
\rule{#1pt}{#2pt}\hskip-#1pt
\rule[#1pt]{#1pt}{#2pt}}\rule[#1pt]{#2pt}{#2pt}\hskip-#2pt
\rule{#2pt}{#1pt}}

\newcommand{\Yfund}{\raisebox{-.5pt}{\drawsquare{6.5}{0.4}}}
\newcommand{\Yasymm}{\raisebox{-3.5pt}{\drawsquare{6.5}{0.4}}\hskip-6.9pt%
        \raisebox{3pt}{\drawsquare{6.5}{0.4}}}

\begin{document}
\baselineskip 0.7cm

\setcounter{footnote}{1}

\begin{titlepage}
\today
\begin{flushright}
UT-826
\end{flushright}

\vskip 0.35cm
\begin{center}
  {\large \bf Branes and Vector-like Supersymmetry Breaking Theories\\
    with Gauged Global Symmetry} 
  \vskip 1.2cm Jiro Hashiba
 \vskip 0.4cm

  {\it Department of Physics, University of Tokyo, Tokyo 113, Japan}

  \vskip 1.5cm
  
  \abstract{We show that the brane configuration describing the
    Izawa-Yanagida-Intriligator-Thomas (IYIT) model with gauged
    $U(1)$ subgroup of the global symmetry contains inconsistent
    geometry, implying that there exists a stable vacuum
    where supersymmetry is dynamically broken.}
\end{center}
\end{titlepage}

\renewcommand{\theequation}{\thesection.\arabic{equation}}
\renewcommand{\thefootnote}{\arabic{footnote}}
\setcounter{footnote}{0}

%
%
%
%

\section{Introduction}
\setcounter{equation}{0}

Dynamical supersymmetry breaking (DSB) is a longstanding problem in
supersymmetric gauge theories \cite{PT}. It provides a natural solution to
the hierarchy problem that there is a large difference in magnitude
between the electroweak scale and the Planck scale, by generating
dynamically strong coupling scales which are far below the Planck scale
(or the GUT scale) because of a logarithmic running of gauge couplings.
There are many known examples of the theories which realize DSB
as consequences of strong coupling dynamics, such as non-pertubatively
generated superpotential \cite{ADS}, quantum deformation of moduli space
\cite{IYIT}, confinement \cite{ISS}, and so on. It is worth noting that
most of the DSB models discovered recently depend on a series of the
exact results in supersymmetric gauge theories, which are obtained by
holomorphy and symmetry arguments \cite{IS1,IS2,IP}. This tells us that
if a new tool for investigating supersymmetric gauge theories is
introduced, it will be a valuable attempt to apply it to
the problem of determining if DSB occurs or not.

In recent developments in string theories and M-theory, the worldvolume
dynamics on various branes has been proved to be a powerful alternative
to analyzing supersymmetric gauge theories \cite{GK}. 
Up to now, it has been found that many of the field theories with
different number of supersymmetries can be realized on branes, and
the properties of the theories which were understood using purely
field theoretical methods so far can be derived from the brane analyses.
For instance, we can construct $N=4$ super Yang-Mills theories on
parallel D3-branes in type IIB string theory, with Montonen-Olive
duality originating from type IIB $SL(2,{\bf Z})$ duality
\cite{Tseytlin,GG}. $N=2$ supersymmetric QCD is realized by connecting
suitably D4-branes and NS5-branes in type IIA string theory, or by
reinterpreting them as describing a single 5-brane in M-theory to take
quantum effects into account
\cite{Witten1,BSTY1,BSTY2,LLL1,AOT1,Terashima,LLL2}. One of the most
remarkable consequences is that the Riemann surface which determines
exactly the metric on Coulomb branch is a two dimensional part of the
M5-brane, with the remaining four dimensional worldvolume being the flat
space-time.

Brane construction of the theories with $N=1$ supersymmetry will be
most interesting for phenomenological applications. Non-perturbative
aspects of field theories, such as runaway behaviour due to
Afleck-Dine-Seiberg type superpotential, quantum modification of
moduli spaces, etc. are all rephrased in M5-brane language
\cite{Witten2,HOO,AOT1,BHOO,AOT2}. Furthermore, we can argue Seiberg's
$N=1$ duality to be nothing but the exchange of two NS5-branes in type
IIA configuration, or the smooth deformation of an M5-brane
\cite{EGK,EJS,BH,BSTY1,EGKRS,CS,Sugimoto,Hori1}. Given that
the correspondence between field theories and branes was established in
this way, we are naturally led to a question as mentioned in the first
paragraph of this section: Can brane configuration encode the
information about DSB?

The authors of \cite{BHOO} gave an answer to this question, by considering
the brane configuration describing the vector-like gauge theory
which breaks supersymmetry dynamically, proposed by Izawa, Yanagida and
Intriligator, Thomas \cite{IYIT} (we refer to this theory as the IYIT
model in the following). In \cite{Hori2} it was argued that
``t-configuration'', the transverse intersection of an M5-brane with an
orbifold fixed plane, is incompatible with Dirac's quantization
condition for 4-form field strength in M-theory, and the M5-brane
configuration for the IYIT model unavoidably contains this
t-configuration, triggering DSB. Therefore, if we admit the
interpretation of t-configuration in M5-brane as DSB in field theory, we
may expect that certain field theories can be shown to have
non-supersymmetric stable vacua by proving that t-configuration appears
inevitably in the corresponding M5-brane configuration.

In this paper we adopt this brane approach to the problem of DSB
in order to investigate the IYIT model with gauged subgroup of
the flavour symmetry. Our motivation to consider this generalization
of the original IYIT model comes from gauge-mediated supersymmetry
breaking \cite{GR,gm,INTY}.
Most of the gauge-mediated supersymmetry breaking models consist of
three sectors, the first one to break supersymmetry dynamically
(hidden sector), the second one to mediate the effects of the broken
supersymmetry (messenger sector), and the last one which governs the
dynamics of the ordinary particles (observable sector).
Generally, we gauge the $U(1)$ subgroup of the global symmetry
possessed by the model exhibiting DSB which serves as the hidden 
sector, so that DSB effects should be transmitted to the messenger sector
through the gauged $U(1)$ interaction. We conclude making use of the
brane technique that the supersymmetry breaking stable vacuum still
remains even if $U(1)$ subgroup of the global symmetry is gauged. This
agrees with the naive expectation from the field theory that gauging
$U(1)$ symmetry does not alter the asymptotic behaviour of the potential 
in the original model.

The organization of this paper is as follows. In section 2,
we briefly review the original IYIT model in both of field theory and
brane context. In section 3, we discuss electric-magnetic duality in $Sp$
gauge theories with gauged global symmetry. In section 4, we demonstrate
explicitly that the M5-brane configuration for the case of
gauged $U(1)$ subgroup of the global symmetry involves t-configuration,
which is a signal of the non-supersymmetric stable vacuum.
The last section is devoted to conclusions.

\section{Review of the Izawa-Yanagida-Intriligator-Thomas Model}
\setcounter{equation}{0}

\subsection{The Original Model}

The theory is based on $N=1$ $Sp(\tilde{N}_c)$ SQCD with $2N_{f}$ quarks
$\tilde{Q}_i^{\tilde{a}}$ and $N_f(2N_f-1)$ gauge singlets $S^{ij}=-S^{ji}$
($\tilde{a},\tilde{b}=1,2,\cdots ,2\tilde{N}_c;i,j=1,2,\cdots ,2N_f$),
with tree-level superpotential
\footnote{
Since it is convenient to consider these theories as magnetic dual
descriptions of electric theories in the next subsection, we interpret
the singlets $S^{ij}$ as mesons of dimension two, and introduce mass
scale $\mu$ in (\ref{IYITsuperpot}).
}
\begin{equation}
  \label{IYITsuperpot}
  \tilde{W}_{tree}=\frac{1}{2\mu}S^{ij}\tilde{Q}_{i}\tilde{Q}_{j},
\end{equation}
where $\tilde{Q}_{i}\tilde{Q}_{j}$ denotes
$J_{\tilde{a}\tilde{b}}\tilde{Q}_i^{\tilde{a}}\tilde{Q}_j^{\tilde{b}}$,
and $J_{\tilde{a}\tilde{b}} = ({\bf 1}_{\tilde{N}_c}\otimes
i\sigma_2)_{\tilde{a}\tilde{b}}$ is the antisymmetric invariant tensor
of $Sp(\tilde{N}_c)$. The representation of the quarks and the singlets
under the gauge and the anomaly free global symmetry $SU(2N_{f}) \times
U(1)_{R}$ is
\begin{equation}
\label{IYITfield}
\begin{array}{c|c|cc}&
Sp(\tilde{N}_{c}) & SU(2N_{f}) & U(1)_R \\ \hline
\tilde{Q} & \Yfund & \overline{\Yfund} & 1-\frac{\tilde{N}_{c}+1}{N_{f}}  \\
S & 1 & \Yasymm & 2\frac{\tilde{N}_{c}+1}{N_{f}}
\end{array}
\end{equation}
The IYIT model corresponds to the case $N_{f}=\tilde{N}_{c}+1$ \cite{IYIT}.
Under this circumstance no non-perturbative superpotential is generated,
because the tree-level superpotential (\ref{IYITsuperpot}) itself is
the only possible one we can write down taking into account holomorphy
and the symmetries of the theory. Instead of generating non-perturbative
superpotential, however, strong coupling effects modify
the classical constraint ${\rm Pf}(\tilde{Q}_i\tilde{Q}_j)=0$
to the quantum one \cite{IP}
\begin{equation}
\label{qconstraint}
  {\rm Pf}(\tilde{Q}_{i}\tilde{Q}_{j}) = \tilde{\Lambda}^{2(\tilde{N}_{c}+1)},
\end{equation}
where $\tilde{\Lambda}$ is the scale at which the gauge coupling
blows up. Since the $F$ term condition for $S^{ij}$,
$\partial W/\partial S^{ij} = 0$, clearly contradicts the constraint
(\ref{qconstraint}), supersymmetry is spontaneously broken.

We can ascertain that supersymmetry is indeed broken from another point
of view, by considering large values for singlets,
$\mu^{-1}S^{ij} \gg \tilde{\Lambda}$. Since all the quarks acquire large
masses under this condition, the low energy effective theory is
described by pure $Sp(\tilde{N}_{c})$ SYM with the scale
$\tilde{\Lambda}_L$ which is related to $\tilde{\Lambda}$ by the coupling
matching, $\tilde{\Lambda}_L^{3(\tilde{N}_{c}+1)}
={\rm Pf}(\mu^{-1}S^{ij})\tilde{\Lambda}^{2(\tilde{N}_{c}+1)}$.
Then, gaugino condensation generates non-pertubative superpotential
\begin{equation}
  \label{condensation}
  \tilde{W}_{eff}=(\tilde{N}_{c}+1)\tilde{\Lambda}_L^3=(\tilde{N}_{c}+1)
       [{\rm Pf}(\mu^{-1}S^{ij})]^{1/(\tilde{N}_{c}+1)}\tilde{\Lambda}^2.
\end{equation}
Since the superpotential (\ref{condensation}) is linear in $S$ (recall
$N_{f}=\tilde{N}_{c}+1$), $F$ term condition $\partial W_{eff}/\partial
S = 0$ cannot be satisfied, leading to DSB. Computation of the one-loop
correction to the K\"{a}hler potential for $S$ reveals that the scalar
potential for $S$ increases in the region $\mu^{-1}S^{ij} \gg
\tilde{\Lambda}$ where perturbative calculation is reliable. Therefore,
we can assert that supersymmetry breaking stable vacuum exists, although
its location in field space cannot be determined because strong coupling
effects on the K\"{a}hler potential for small $S$ is uncontrollable.

\subsection{Brane Analysis}

We can regard $N=1$ $Sp(\tilde{N}_c)$ gauge theory with matter
representation (\ref{IYITfield}) and tree-level superpotential
(\ref{IYITsuperpot}) as a magnetic dual description of
$Sp(N_c(=N_f-\tilde{N}_c-2))$ SQCD with $2N_f$ massless quarks,
provided that the number of flavours satisfies $N_f>\tilde{N}_c +2$.
If we put $N_f=\tilde{N}_c +1$ ignoring this inequality,
the IYIT model can be obtained as electric $Sp(N_c(=-1))$ SQCD.
In \cite{BHOO}, the IYIT model was analyzed by constructing the M5-brane
configuration for $Sp(-1)$ gauge theory, as we review in this subsection.

In order to realize $Sp$ gauge theory with $N=1$ supersymmetry
on the worldvolume of the branes in type IIA string theory, we require
the following branes and orientifold 4-plane (O4-plane),
\begin{equation}
\label{IIAbranes}
\begin{array}{lcl}
    {\rm NS5} &:& 012345  \\
   {\rm NS'5} &:& 012389  \\
     {\rm D6} &:& 0123789  \\
     {\rm D4} &:& 01236  \\
     {\rm O4} &:& 01236
\end{array}
\end{equation}
where the numbers indicate the directions in the ten dimensional flat
space-time along which the branes and the O4-plane are extending
\footnote{
We use the terminology ``NS5-brane'' in two meanings, the one refers to
general NS5-branes with their worldvolume extending in arbitray
directions including NS$'$5-brane in (\ref{IIAbranes}), and the other to 
the NS5-branes of the type specified in (\ref{IIAbranes}).
}.
It needs mentioning the peculiar nature of the O4-plane in the presence of
the other branes in (\ref{IIAbranes}). If the O4-plane exists, we must
identify the two points which are related to each other by
the ${\bf Z}_2$ transformation $x^{4,5,7,8,9}\rightarrow -x^{4,5,7,8,9}$.
There are two kinds of O4-plane distinguished by their Ramond-Ramond
(RR) charges, namely the O4-planes with RR charge $+1$ and $-1$ which realize
$Sp$ and $SO$ gauge theory respectively on the worldvolume of D4-branes
which are set parallel to the O4-plane so as to obey the ${\bf Z}_2$
symmetry \cite{EJS}. When an NS5-brane intersects with an O4-plane so
that they share a real codimension one subspace of the O4-plane,
the two regions into which the O4-plane is divided by the NS5-brane have
opposite RR charges, as if the NS5-brane behaves as a domain wall \cite{EGKT}.

The type IIA brane configuration for $N=1$ $Sp(N_c)$ SQCD with $2N_f$
massless quarks can be constructed as follows. We introduce the complex
coordinates $v$ and $w$,
\begin{equation}
  \left\{
       \begin{array}{l}
          v=x^4+ix^5, \\
          w=x^8+ix^9.
       \end{array} \right.
\end{equation}
We first put an O4-plane
at $v=0,w=0$
\footnote{
Throughout this paper, all the branes except the D6-branes lie on the
hyperplane $x^7=0$.
},
and place an NS5-brane at $w=0$ and an NS$'$5-brane at $v=0$, with their 
positions in $x^6$ direction different. Here, we choose the RR charge
carried by the O4-plane between the NS5- and the NS$'$5-brane to be
$+1$, in order to guarantee the gauge group to be $Sp(N_c)$. Then, we
connect the two NS5-branes by $2N_c$ D4-branes, and locate $2N_f$
D6-branes at $v=0$ and arbitrary $x^6$ positions on the interval between
the two NS5-branes. The brane configuration made this way is shown in Figure 1.
The $Sp(N_c)$ vector multiplet corresponds to the string both ends of
which are stuck on the D4-branes, and $2N_f$ quarks are identified with
the strings linking the D4-branes and the D6-branes.
\begin{figure}
    \centerline{\psfig{figure=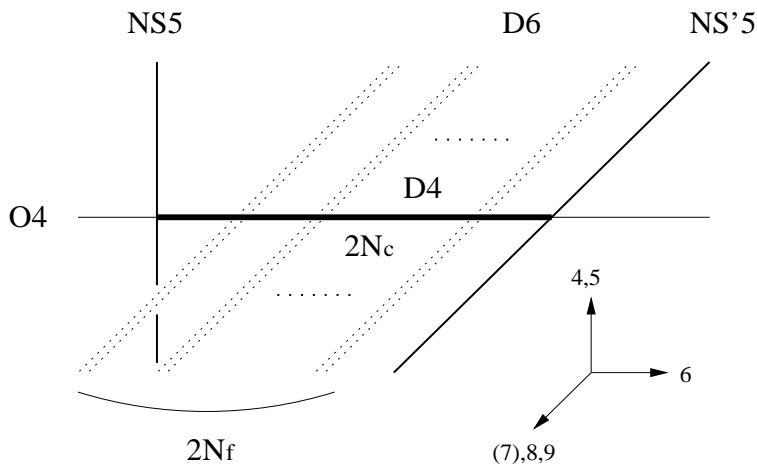,width=10cm}}
    \caption{Type IIA brane configuration for the electric $Sp(N_c)$ SQCD.}
\end{figure}

Before lifting the above electric type IIA configuration up to a single
M5-brane, we present here the type IIA brane configuration
for the magnetic theory which expresses the IYIT model under the
condition $N_{f}= \tilde{N}_{c}+1$. The duality transformation
in field theory is equivalent in the brane picture
to moving the NS5-brane to the right of the NS$'$5-brane\cite{EJS}.
First, the NS5-brane passes through the $2N_{f}$ D6-branes,
with a D4-brane created between the NS5-brane and a D6-brane
just at the moment when the NS5-brane and the D6-brane coincide in the
$x^{6}$ direction. At this stage we have $2N_{f}$ D4-branes,
each of which is suspended between the NS5-brane and one of the
$2N_{f}$ D6-branes. Then, we connect $2N_{c}$ D4-branes between the
NS5-brane and the NS$'$5-brane to $2N_{c}$ of the $2N_{f}$ D4-branes
ending on the $2N_{f}$ D6-branes, so that $2N_{f}-2N_{c}$ D4-branes
should connect the NS5-brane and $2N_{f}-2N_{c}$ D6-branes. Finally we let
the NS5-brane pass through the NS$'$ 5-brane. When the NS5-brane and the
right NS$'$5-brane collide, two of the D4-branes and their mirror
partners between the NS5-brane and D6-branes disappear \cite{EJS},
leading to $2\tilde{N_{c}}=2(N_{f}-N_{c}-2)$ D4-branes suspended between
the NS$'$5-brane and the NS5-brane. We reach eventually the brane
configuation shown in Figure 2.
\begin{figure}
    \centerline{\psfig{figure=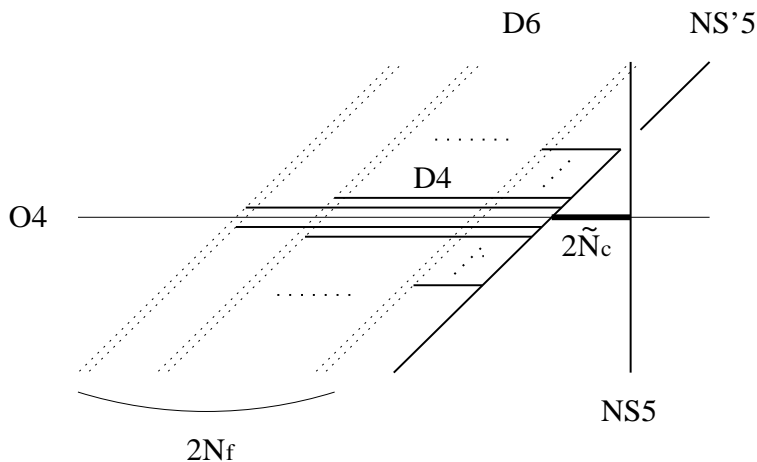,width=10cm}}
    \caption{Type IIA brane configuration for the magnetic $Sp(\tilde{N}_c)$ SQCD.}
\end{figure}

We now discuss how the type IIA brane configurations given above are
embedded in M-theory, by introducing an extra dimension with coordinate
$x^{10}$, which is a circle of radius $R$. The O4-plane with RR charge
$-1$ originates from the five dimensional orbifold fixed plane whose
worldvolume is extending in the direction $x^{0,1,2,3,6,10}$, while the
M-theory counterpart of the O4-plane with RR charge $+1$ is the fixed
plane screened by two M5-branes on top of it. 

The D6-branes are interpreted as Kaluza-Klein monopoles, the four
dimensional space transverse to which is described by the Taub-NUT
geometry. Although the Taub-NUT space possesses hyperk\"{a}hler
structure, it suffices to choose one complex structure of that, for the
purpose of embedding an M5-brane. In the present paper we express the
Taub-NUT space by a two complex dimensional hypersurface in ${\bf C}^3$
with complex coordinates $(x,y,v)$,
\begin{equation}
\label{Taub-NUT}
   xy=v^{2N_f}.
\end{equation}
It is reflected on the $A_{2N_f-1}$ type singularity at the origin
$x=y=v=0$ that the D6-branes are located at the same position in the $v$
direction, $v=0$. However, the singularity must be resolved in a generic
situation where the $x^6$ positions of the D6-branes are different,
to describe correctly the geometry around the D6-branes \cite{Witten1}.
The resolved surface is covered by $2N_f$ patches $U_i~(i=1,2,\cdots
,2N_f)$ with coordinates $(x_i,y_i)$, which are glued by the transformation
$x_i y_{i+1}=1,x_i y_i=x_{i+1}y_{i+1}$. The three cordinates $(x,y,v)$ are
related to the $U_i$ coordinates by
\begin{equation}
\label{Taub-NUT-map}
   x=x_i^{2N_f-i+1}y_i^{2N_f-i},~y=x_i^{i-1}y_i^i,~v=x_i y_i.
\end{equation}
Upon resolving the singularity there appear $2N_f-1$ exceptional ${\bf
P}^1$ cycles $C_i~(i=1,2,\cdots ,2N_f-1)$ which are the loci $y_i=0$ in 
$U_i$ or $x_{i+1}=0$ in $U_{i+1}$, with two adjacent components
intersecting transversely with one another at the position of a
D6-brane. Under the ${\bf Z}_2$ symmetry the coordinates on $U_i$
transform as $(x_i,y_i)\rightarrow((-1)^i x_i,(-1)^{i+1}y_i)$, implying
that $C_i$ with even $i$ and the two infinite planes $C_y$ and $C_x$
defined by $x_1=0$ and $y_{2N_f}=0$ respectively are ${\bf Z}_2$ invariant,
while $C_i$ with odd $i$ are ${\bf Z}_2$ reversed. The orbifold fixed
plane is thus divided into the two infinite planes $C_y$, $C_x$ and
the $N_f-1$ ${\bf P}^1$'s, i.e. $C_i$ with even $i$. We depict the
Taub-NUT space schematically in Figure 3.
\begin{figure}
    \centerline{\psfig{figure=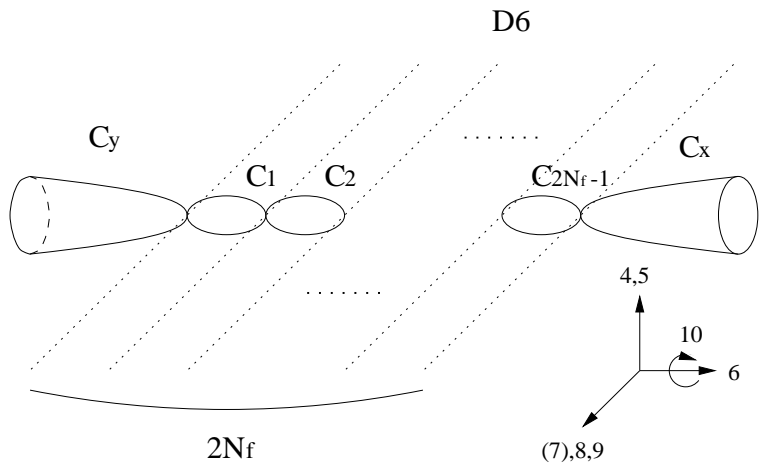,width=10cm}}
    \caption{The Taub-NUT space.}
\end{figure}

To summarize, the structure of the eleven dimensional space-time is
${\bf R}^4 \times ({\bf M}^7/{\bf Z}_2)$, where ${\bf R}^4$ is the ordinary 
four dimensional space-time, ${\bf M}^7$ the product of the Taub-NUT space and 
the ${\bf R}^3$ parametrized by $x^{7,8,9}$, and the ${\bf Z}_2$ acts
on ${\bf M}^7$ as $(x,y,v,w,x^7) \rightarrow (x,y,-v,-w,-x^7)$.

On this manifold we embed a single M5-brane into which the D4-branes and 
the NS5-branes combine. The M5-brane worldvolume is of the form
${\bf R}^4 \times \Sigma$, where $\Sigma$ is a Riemann surface in
${\bf M}^7/{\bf Z}_2$. Our general procedure to extract the information on the
vacuum structure of a supersymmetric field theory from the Riemann
surface $\Sigma$ is as follows \cite{BHOO}. We impose an appropriate boundary
condition on the M5-brane which is holomorphic with respect to the
complex structure of the manifold (non-holomorphic boundary
conditions are imposed instead, when we consider the theories with
supersymmetry {\it explicitly} broken). The M5-brane configurations which are
holomorphic everywhere and satisfy the boundary condition ensure
supersymmetry preserving vacua to exist. Generically, the holomorphic
configurations are parametrized by a number of parameters, which
correspond to the moduli space in the field theory. On the other hand, it is
possible that no M5-brane can fulfill the holomorphy and the boundary
condition simultaneously. This situation indicates that supersymmetry is
{\it spontaneously} broken, even though we need more information to
determine whether the field theory exhibits runaway behaviour or the
potential has local minima where supersymmetry is dynamically broken.

As a result of applying the procedure explained above to the
$Sp(-1)$ gauge theory, we obtain the unique {\it holomorphic} M5-brane
configuration which consists of two components,
\begin{equation}
  \label{IYITconf}
    C_L \left\{
            \begin{array}{l}
              y=1, \\
              w=0,
            \end{array} \right. \hspace{2cm}
    C_R \left\{
            \begin{array}{l}
              x=(\mu^{-1}\tilde{\Lambda}^2)^{2N_f}, \\
              v=0,
            \end{array} \right.
\end{equation}
where $C_L$ and $C_R$ reduce to the NS5-brane and the NS$'$5-brane in
Figure 1 in a suitable type IIA limit. One might wonder how these
everywhere holomorphic curves are consistent with DSB in the IYIT model.
In order to solve this discrepancy, note that the component $C_L$
transversely intersects with the infinite fixed plane $C_y$, sharing
${\bf R}^4$ part of their worldvolumes
\footnote{
In the following, the ``transverse'' intersection between an NS5-brane
and the orbifold fixed plane always means this type of intersection.
}
($C_R$ is intersecting with $C_x$
in the same manner). The rules of possible intersections between
M5-branes and orbifold fixed planes are obtained in \cite{Hori2}, by examining
the flux quantization condition for the 4-form field strength in
M-theory. According to the rules, the transverse intersection between a
single M5-brane and an infinite fixed plane is forbidden, while there
is no obstruction to the same type of intersection if the infinite fixed
plane is screened by a pair of M5-brane. Therefore, supersymmetry is
spontaneously broken because the unique holomorphic configuration
(\ref{IYITconf}) is not allowed. The non-holomorphic M5-brane
configuration corresponding to the supersymmetry breaking stable vacuum
is expected to exist, although we cannot make any definite statements
about its exact form or the vacuum energy, due to the breakdown of the
supergravity approximation of M-theory in the vicinity of the orbifold
fixed plane.

\section{Electric-Magnetic Duality in SQCD with Gauged Global Symmetry}
\setcounter{equation}{0}

In this section we study the electric-magnetic duality in $N=1$ $Sp(N_{c})$
Supersymmetric QCD in the case where $SO(2N_{c}')$ subgroup of the
global symmetry is gauged. This duality is useful when we discuss
dynamically broken supersymmetry using the technique of M5-brane in the
next section, since the M5-brane configuration is more easily obtained from
the type IIA brane configuration of the electric theory than from
that of the magnetic one, which describes the IYIT model with
gauged global symmetry if we put formally $N_{c}=-1$ as in the previous 
section. Dualities in supersymmetric QCD with product gauge groups are
also discussed in \cite{PST} from the ordinary field theory point
of view, and in \cite{BSTY1,AOT3,LO} by means of brane configurations.

The matter contents in the electric theory are $2N_{f}$ quarks
$Q_a^i$ in the fundamental representation under $Sp(N_{c})$ and a quark
$q_a^s$ in the bifundamental representation under
$Sp(N_{c}) \times SO(2N_{c}')$ $(a,b=1,2,\cdots ,2N_c;s,t=1,2,\cdots
,2N_c')$. The anomaly free global symmetry is
$SU(2N_{f}) \times U(1)_{R}$, and the representation of the quarks under
the gauge and the global symmetries are summarized as follows.
\begin{equation}
\label{efield}
\begin{array}{c|cc|cc}
&Sp(N_{c}) & SO(2N_{c}') & SU(2N_{f}) & U(1)_R \\ \hline
Q & \Yfund & 1 & \Yfund & 1+\frac{N_{c}'(N_{c}'-1)}{N_{f}N_{c}}
                          -\frac{N_{c}+1}{N_{f}}  \\
q & \Yfund & \Yfund & 1 & 1-\frac{N_{c}'-1}{N_{c}}.
\end{array}
\end{equation}

For $N_{f}+N_c' > N_c+2$, we suspect that this electric theory has a
dual magnetic description with gauge group $Sp(\tilde{N}_c) \times
SO(2N_c')$ where $\tilde{N}_c=N_{f}+N_c'-N_c-2$, regarding $SO(2N_c')$
gauge group as a spectator. The transformation properties of magnetic
fields under the gauge and the global symmetries are
\begin{equation}
\label{mfield}
\begin{array}{c|cc|cc}
&Sp(\tilde{N}_{c}) & SO(2N_{c}') & SU(2N_{f}) & U(1)_R \\ \hline
\tilde{Q} & \Yfund & 1 & \overline{\Yfund}
          & -\frac{N_{c}'(N_{c}'-1)}{N_{f}N_{c}}+\frac{N_{c}+1}{N_{f}}  \\
\tilde{q} & \Yfund & \Yfund & 1
          & \frac{N_{c}'-1}{N_{c}}  \\
\Phi & 1 & \Yasymm & 1
          & 2-2\frac{N_{c}'-1}{N_{c}}  \\
N & 1 & \Yfund & \Yfund
          & 2+\frac{N_{c}'(N_{c}'-1)}{N_{f}N_{c}}-\frac{N_{c}+1}{N_{f}}
            -\frac{N_{c}'-1}{N_{c}}  \\
S & 1 & 1 & \Yasymm
          & 2+2\frac{N_{c}'(N_{c}'-1)}{N_{f}N_{c}}-2\frac{N_{c}+1}{N_{f}},
\end{array}
\end{equation}
where $\tilde{Q}_i^{\tilde{a}}$ and $\tilde{q}_s^{\tilde{a}}$ are
magnetic quarks, and $\Phi^{st}$, $N^{is}$ and $S^{ij}$ are ``mesons''
described in terms of the electric fields as $\Phi^{st}=q^s q^t$,
$N^{is}=Q^i q^s$ and $S^{ij}=Q^i Q^j$. In order for the two theories to
flow into the same fixed point in the IR limit, we must also incorporate
the following tree-level superpotential:
\begin{equation}
\label{msuperpot}
  \tilde{W}_{tree}=\frac{1}{2\mu}S^{ij}\tilde{Q}_{i}\tilde{Q}_{j}
          +\frac{1}{2\mu}\Phi^{st}\tilde{q}_{s}\tilde{q}_{t}
          +\frac{1}{\mu}N^{is}\tilde{Q}_{i}\tilde{q}_{s}.
\end{equation}
The dimensionful parameter $\mu$ in (\ref{msuperpot}) relates
the two scales $\Lambda,\Lambda'$ for $Sp(N_c),SO(2N_c')$ in the
electric theory to the scales $\tilde{\Lambda},\tilde{\Lambda}'$ for
$Sp(\tilde{N_{c}}),SO(2N_c')$ in the magnetic theory as
\footnote{
In this section, we omit the irrelevant numerical factors from the scale 
mathing relations.
}
\begin{equation}
  \label{Lambda}
  \Lambda^{3(N_c+1)-(N_f+N_c')}\tilde{\Lambda}^{3(\tilde{N}_c+1)-(N_f+N_c')}
     = (-1)^{N_f+N_c'-N_c-1}\mu^{N_f+N_c'},
\end{equation}
\begin{equation}
  \label{Lambda'}
  \Lambda^{6(N_c+1)-2(N_f+N_c')}\Lambda^{\prime 6(N_c'-1)-2N_c}
     = \mu^{2(N_f+N_c+N_c')}
       \tilde{\Lambda}^{\prime 4(N_c'-1)-2(N_f+\tilde{N}_c)}.
\end{equation}

A non-trivial consistency check of the duality is provided by 't Hooft
anomaly matching condition. For appropriate choices of $N_f$, $N_c$, and
$N_c'$, we expect that the dual theories have a common moduli space with
the origin where the full global symmetry is left unbroken. At least in 
such a situation, we can verify that the global anomalies match at the
origin, from the charge assignments to the fermions in the two theories:
\begin{equation}
\label{anomaly}
\begin{array}{rl}
SU(2N_{f})^3 & 2N_c d^{(3)}(2N_f), \\
SU(2N_{f})^2 U(1)_R & 2\frac{N_c'(N_c'-1)-N_c(N_c+1)}{N_f}d^{(2)}(2N_f), \\
U(1)_R^3 & N_c(2N_c+1)+N_c'(2N_c'-1)
           +4\frac{[N_c'(N_c'-1)-N_c(N_c+1)]^3}{N_f^2N_c^2}
           -4\frac{N_c'(N_c'-1)^3}{N_c^2}, \\
U(1)_R & -N_c(2N_c+3)+N_c'(2N_c'-1).
\end{array}
\end{equation}

In the remainder of this section, we consider various decoupling limits of
the two theories to give further evidence for the duality, and in
particular to confirm the coupling matching relations (\ref{Lambda}) and
(\ref{Lambda'}).

\bigskip
{\it Mass Deformations for $Q$}

Let us start with deforming the electric theory by adding a mass term
for $Q^{2N_f-1}$ and $Q^{2N_f}$, $W_{tree}=mQ^{2N_f-1}Q^{2N_f}$.
At an energy scale below the quark mass $m$, 
$Sp(N_c) \times SO(2N_c')$ gauge group is unbroken and the matter
spectrum consists of $2N_f-2$ quarks $Q^{\hat{\imath}}$ and one bifundamental
$q$, where $\hat{\imath}$ runs over $N_f-1$ light flavours,
$\hat{\imath}=1,2,\cdots,2N_f-2$. The dynamical scale for $Sp(N_c)$ in
the low energy effective theory $\Lambda_L$ is thus determined by the relation
\begin{equation}
  \label{scale1}
  \Lambda_L^{3(N_c+1)-(N_f-1+N_c')}=m\Lambda^{3(N_c+1)-(N_f+N_c')},
\end{equation}
while $\Lambda'$ remains unchanged. The tree-level superpotential in the 
magnetic theory takes the form
\begin{equation}
\label{msuperpot-mass}
  \tilde{W}_{tree}=\frac{1}{2\mu}S^{ij}\tilde{Q}_i\tilde{Q}_j
          +mS^{2N_f-1,2N_f}
          +\frac{1}{2\mu}\Phi^{st}\tilde{q}_s\tilde{q}_t
          +\frac{1}{\mu}N^{is}\tilde{Q}_i\tilde{q}_s.
\end{equation}
$F$ term condition for $S^{2N_f-1,2N_f}$ is
$\langle \tilde{Q}_{2N_f-1}\tilde{Q}_{2N_f} \rangle = -\mu m$, which
means that by an appropriate gauge ans globalrotation we can put
$\langle \tilde{Q}^{\tilde{a}}_{2N_f-1} \rangle =
(-\mu m)^{1/2}\delta_{2N_c-1}^{\tilde{a}},
\langle \tilde{Q}^{\tilde{a}}_{2N_f} \rangle =
(-\mu m)^{1/2}\delta_{2N_c}^{\tilde{a}}$
with the vev's of the other quarks vanishing. These vev's break
$Sp(\tilde{N}_c)$ to $Sp(\tilde{N}_c-1)$ and give rise to the mass squared
$-\mu m$ of $S^{\hat{\imath},2N_f-1},S^{\hat{\imath},2N_f},S^{2N_f-1,2N_f}$ and
$\tilde{Q}_{2N_f-1},\tilde{Q}_{2N_f}$. Matching $Sp(\tilde{N}_c)$
coupling at the scale $-\mu m$, we obtain $\tilde{\Lambda}_L$, the scale
for $Sp(\tilde{N}_c)$ in the low energy theory, as
\begin{equation}
  \label{scale2}
  \tilde{\Lambda}_L^{3(\tilde{N}_c-1+1)-(N_f-1+N_c')}
      =-(\mu m)^{-1}\tilde{\Lambda}^{3(N_c+1)-(N_f+N_c')}.
\end{equation}
From eqs. (\ref{scale1}) and (\ref{scale2}) we reproduce eq. (\ref{Lambda})
with $N_f$ and $\tilde{N}_c$ replaced by $N_f-1$ and $\tilde{N}_c-1$.
Due to the quark vev's, $SO(2N_c')$ fundamentals
$\tilde{q}^{2\tilde{N}_c-1},\tilde{q}^{2\tilde{N}_c}$ and
$N^{2N_f-1},N^{2N_f}$ also acquire the mass squared $-\mu m$, leading to 
the relation between $\tilde{\Lambda}'$ and the scale
$\tilde{\Lambda}'_L$ in the low energy theory
\begin{equation}
  \label{scale3}
  \tilde{\Lambda}_L^{\prime 4(N_c'-1)-2(N_f-1+\tilde{N}_c-1)}
      =(-\mu m)^2\tilde{\Lambda}^{\prime 4(N_c'-1)-2(N_f+\tilde{N}_c)}.
\end{equation}
Using eqs. (\ref{scale1}) and (\ref{scale3}), we find that the scales in 
the low energy effective theory satisfy eq. (\ref{Lambda'}).

\bigskip
{\it Flat Directions for $Q$}

We can analyze another decoupling limit by the vev's of the electric
quarks $Q$ along their $D$-flat directions. First examine the case
where $\langle Q^{2N_f-1}Q^{2N_f} \rangle = a^2$ and all the other meson 
vev's vanish. Below the scale $a$ the electric theory flows to
$Sp(N_c-1) \times SO(2N_c')$ SQCD with $2N_f-2$ quarks
$Q_{\hat{a}}^{\hat{\imath}}$, one bifundamental $q_{\hat{a}}^s$ (here,
$\hat{a}$ refers to $N_c-1$ colors, $\hat{a}=1,2,\cdots,2N_c-2$) and two
$SO(2N_c')$ fundamental quarks $q_{2N_c-1},q_{2N_c}$ with the dynamical
scale $\Lambda_L$ for $Sp(N_c-1)$ given by
\begin{equation}
  \label{scale4}
  \Lambda_L^{3(N_c-1+1)-(N_f-1+N_c')}=a^{-2}\Lambda^{3(N_c+1)-(N_f+N_c')}.
\end{equation}
In the magnetic theory, $\langle S^{2N_f-1,2N_f} \rangle = a^2$ induces
the mass $\mu^{-1}a^2$ of $\tilde{Q}_{2N_f-1}$ and $\tilde{Q}_{2N_f}$.
Therefore, the dynamical scale $\tilde{\Lambda}_L$ for $Sp(\tilde{N}_c)$
in the low energy magnetic theory is related to $\tilde{\Lambda}$ by
\begin{equation}
  \label{scale5}
  \tilde{\Lambda}_L^{3(\tilde{N}_c+1)-(N_f-1+N_c')}
    =\mu^{-1}a^2\tilde{\Lambda}^{3(\tilde{N}_c+1)-(N_f+N_c')},
\end{equation}
which, combined with (\ref{scale4}), shows that the relation
(\ref{Lambda}) is preserved.

On the other hand, eq. (\ref{Lambda'}) is not recovered in the low
energy effective theory, since all the spectra with non-trivial $SO(2N_c')$
representation do not decouple in both of the dual theories. In other
words, the effective dual theories in a low energy scale are not
described by just replacing $N_f$ and $N_c$ by $N_f-1$ and $N_c-1$
in the high energy dual theories. This fact is, for instance, reflected
on the electric quarks $q_{2N_c-1},q_{2N_c}$ which are coupled to
the other fields via gauged $SO(2N_c')$ interactions.

As a further comparison of the dual theories, we take one particular
choice of the number of flavours, $N_f=N_c$, and let the electric quarks
$Q$ develop such vev's as to satisfy ${\rm Pf}(Q^i Q^j) \neq 0$. 
Then, in the low energy theory $Sp(N_c)$ gauge group is completely
broken, and remain $2N_c$ quarks $q_a$ in fundamental representation
under $SO(2N_c')$ ($a$ is now flavour index) as light matter. If $N_c
\leq N_c'-2$, strong coupling effects of $SO(2N_c')$ gauge dynamics
generate non-perturbative superpotential
\footnote{
For $N_c = N_c'-2$, there is another phase with no superpotential
generated even non-perturbatively \cite{IS2}.
}
\begin{equation}
  \label{ele-eff}
  W_{eff}=(N_c'-N_c-1)\left[\frac{\Lambda^{\prime 6(N_c'-1)-2N_c}}
          {\det(q_a q_b)}\right]^{\frac{1}{2(N_c'-1)-2N_c}}.
\end{equation}

In the magnetic description, all the $2N_f$ quarks $\tilde{Q}$ acquire
mass due to ${\rm Pf}S^{ij} \neq 0$. Consequently, the number of
$Sp(\tilde{N}_c)$ fundamental quarks $\tilde{q}_s$ in the low energy theory is
$2N_c'=2(\tilde{N}_c+2)$, which indicates that s-confinement of
$Sp(\tilde{N}_c)$ charge occurs \cite{IP}. Using the notation
$\tilde{\Phi}_{st}=\tilde{q}_s \tilde{q}_t$ as ``gauge-invariant'' composite
states of the fundamental quarks $\tilde{q}$, s-confining effective
superpotential (plus the tree level one with decoupled fields
eliminated) is given by
\begin{equation}
  \label{mag-eff1}
  \tilde{W}_{eff}
    =-\frac{{\rm Pf} \tilde{\Phi}}{\tilde{\Lambda}_L^{2\tilde{N}_c+1}}
     +\frac{1}{2\mu}\Phi \tilde{\Phi},
\end{equation}
where $\tilde{\Lambda}_L^{2\tilde{N}_c+1}
={\rm Pf}(\mu^{-1}S)\tilde{\Lambda}^{2\tilde{N}_c+1-N_f}$. At the scale
$\tilde{\Lambda}_L$, the elementary quarks $\tilde{q}$ combine
into mesons $\tilde{\Phi}$, and decouple immediately together with
the $SO(2N_c')$ adjoint $\Phi$ because of the mass term in the r.h.s. of
eq. (\ref{mag-eff1}) (note that $\Phi /\mu$ and
$\tilde{\Phi}/\tilde{\Lambda}_L$ are canonically normalized fields).
Eventually, at an energy scale lower than $\tilde{\Lambda}_L$ we are left with
$N_f=N_c \leq N_c'-2$ flavours of $SO(2N_c')$ fundamentals $N^i$ which give
rise to the non-perturbative superpotential
\begin{equation}
  \label{mag-eff2}
  \tilde{W}_{eff}
    =(N_c'-N_f-1)\left[\frac{\tilde{\Lambda}_L^{\prime 6(N_c'-1)-2N_f}}
     {\det(\mu^{-2}N^i N^j)}\right]^{\frac{1}{2(N_c'-1)-2N_f}},
\end{equation}
where $\tilde{\Lambda}_L^{\prime 6(N_c'-1)-2N_f}
=\tilde{\Lambda}_L^{4\tilde{N}_c+2}
\tilde{\Lambda}^{\prime 4(N_c'-1)-2(N_f+\tilde{N}_c)}$. Applying
the expression $N^{is}=J^{ab}Q_a^i q_b^s$ and the scale matching
relations (\ref{Lambda}) and (\ref{Lambda'}) to (\ref{mag-eff2}), the
effective superpotential in the electric theory (\ref{ele-eff}) is
correctly reproduced.

\bigskip
{\it Flat Directions for $q$}

Here, we consider the flat directions for the bifundamental quark $q$,
focussing our attention on one specific example where $N_c > N_c'$
and $\langle q_a^s \rangle = a\delta_a^s$ hold. In the electric theory,
the vev's for $q$ break the gauge group $Sp(N_c) \times SO(2N_c')$ to
$Sp(N_c-N_c') \times U(N_c')$, leaving in the low energy effective
theory the field contents
\begin{equation}
\label{light-efield}
\begin{array}{c|cc|cc}
    & Sp(N_c-N_c') & U(N_c') & SU(2N_{f}) \\ \hline
  R & \Yfund & 1 & \Yfund  \\
  r & 1 & \Yfund & \Yfund  \\
  \bar{r} & 1 & \overline{\Yfund} & \Yfund  \\
  \Phi_{U} & 1 & adj & 1  \\
\end{array}
\end{equation}
where $Q$ is decomposed into $R$, $r$, and $\bar{r}$, and $\Phi_{U}$
comes from $q$. Note that the effective theory splits into the two
systems which do not interact with one another, namely the fields with
$Sp(N_c-N_c')$ quantum number and the fields with $U(N_c')$ quantum
number. The coupling mathcing at the scale $a$ determines the scale
$\Lambda_L$ for $Sp(N_c-N_c')$ and the scale $\Lambda_{SU}$ for
$SU(N_c')$ factor of $U(N_c')$,
\begin{equation}
  \label{Lambda-L}
  \Lambda_L^{3(N_c-N_c'+1)-N_f}=a^{-2N_c'}\Lambda^{3(N_c+1)-(N_f+N_c')},
\end{equation}
\begin{equation}
  \label{eLambda-SU}
  \Lambda_{SU}^{2N_c'-2N_f}=a^{-4N_c-2N_c'}\Lambda^{6(N_c+1)-2(N_f+N_c')}
     \Lambda^{\prime 6(N_c'-1)-2N_c}.
\end{equation}

Turning on $\langle q_a^s \rangle = a\delta_a^s$ induces $\langle
\Phi^{st} \rangle =a^2({\bf 1}_{N_c'}\otimes i\sigma_2)^{st}$, and the gauged
global symmetry $SO(2N_c')$ is broken to $U(N_c')$ in the magnetic
theory. The low energy magnetic theory is described by the following
light fields
\begin{equation}
\label{light-mfield}
\begin{array}{c|cc|cc}
    & Sp(\tilde{N}_c) & U(N_c') & SU(2N_{f}) \\ \hline
  \tilde{R} & \Yfund & 1 & \overline{\Yfund}  \\
  \tilde{r} & 1 & \Yfund & \Yfund  \\
  \tilde{\bar{r}} & 1 & \overline{\Yfund} & \Yfund  \\
  \tilde{\Phi}_{U} & 1 & adj & 1  \\
  S & 1 & 1 & \Yasymm
\end{array}
\end{equation}
and the tree-level superpotential
\begin{equation}
  \label{lowenergy-tree}
  W=\frac{1}{2\mu}S^{ij}\tilde{R}_{i}\tilde{R}_{j}.
\end{equation}

Similarly to the electric theory, the effective theory of the magnetic
description consists of two parts with irrelevant interaction between
them. $\tilde{R}$ and $S$ with the Yukawa coupling
(\ref{lowenergy-tree}) define $Sp(\tilde{N}_c)$ gauge theory, which is
dual to the $Sp(N_c-N_c')$ effective theory in the electric
description. The other part is the $U(N_c')$ gauge theory with the
matter contents in the same representation as those in the electric
theory, the correspondence between the electric and the magnetic fields
being $\tilde{r}=(\mu^{-1}a)r$, $\tilde{\bar{r}}=(\mu^{-1}a)\bar{r}$,
and $\tilde{\Phi}_U=(\mu^{-1}a)\Phi_U$. The scales $\tilde{\Lambda}_L$
and $\tilde{\Lambda}_{SU}$ for the two factors in $Sp(\tilde{N}_c)
\times U(N_c')$ are related to the ones in the high energy theory as
\begin{equation}
  \label{tilde-Lambda-L}
  \tilde{\Lambda}_L^{3(\tilde{N}_c+1)-N_f}=(\mu^{-1}a^2)^{N_c'}
     \tilde{\Lambda}^{3(\tilde{N}_c+1)-(N_f+N_c')},
\end{equation}
\begin{equation}
  \label{mLambda-SU}
  \tilde{\Lambda}_{SU}^{2N_c'-2N_f}=(\mu^{-1}a^2)^{2N_f-2N_c}
     \tilde{\Lambda}^{\prime 4(N_c'-1)-2(N_f+\tilde{N}_c)}.
\end{equation}
From (\ref{Lambda-L}) and (\ref{tilde-Lambda-L}), we can verify that the 
relation (\ref{Lambda}) is indeed preserved in the low energy dual theories. 
However, the two scales for $SU(N_c')$ in the dual theories must be identical 
only up to a constant in order for the relation (\ref{Lambda'}) to be valid:
\begin{equation}
  \label{Lambda-SU-change}
  \tilde{\Lambda}_{SU}^{2N_c'-2N_f}=(\mu^{-1}a)^{4N_f+2N_c'}
     \Lambda_{SU}^{2N_c'-2N_f}.
\end{equation}

The origin of the multiplicative factor in (\ref{Lambda-SU-change}) is
the difference in the normalization factors of the matter fields in the
electric and the magnetic theories \cite{AM}. Suppose that we have canonically
normalized chiral superfields $\phi_i$ in the representation $R_i$ of a
gauge group $G$. If we rescale $\phi_i=Z_i^{-1/2}\tilde{\phi}_i$ and vary
the cutoff to renormalize $D$-terms so that $\tilde{\phi}_i$ should have
the canonical kinetic term, the original scale for the gauge group $G$,
$\Lambda_G$, is changed to the new one $\tilde{\Lambda}_G$ due to the
anomalous Jacobian
$|{\cal D}(Z_i^{-1/2}\tilde{\phi}_i)/{\cal D}\tilde{\phi}_i|^2 \neq 1$, as
\begin{equation}
  \label{scale-change}
  \tilde{\Lambda}_G^{-b}=\Lambda_G^{-b}\prod_i Z_i^{T(R_i)},
\end{equation}
where $b$ is the coefficient of the one-loop $\beta$ function and $T(R)$ the
Dynkin index of the representation $R$. In the present case,
$G=SU(N_c')$, $b=-2N_c'+2N_f$, $T(\Yfund)=T(\overline{\Yfund})=\frac{1}{2}$,
$T(adj)=N_c'$ and $Z_i=(\mu^{-1}a)^2$ for all the matter. Substituting
these in (\ref{scale-change}) yields (\ref{Lambda-SU-change}).

\section{Brane Realization of Dynamical Supersymmetry Breaking}
\setcounter{equation}{0}

In this section we gauge $SO(2N_{c}')~(1 \leq N_{c}' \leq
\tilde{N}_c)$ subgroup of the flavour group $SU(2\tilde{N}_{c}+2)$ in the 
IYIT model, and investigate the resulting theories by the brane
method. In \cite{BHOO} was analyzed the case in which an maximal
subgroup $SO(2\tilde{N}_{c}+2) \subset SU(2\tilde{N}_{c}+2)$ is gauged, the
brane prediction agreeing with the field theory that the model exhibits
runaway behaviour.

We can obtain the type IIA brane configuration realizing
the IYIT model with gauged $SO(2N_{c}')$ subgroup of the global symmetry
(the magnetic description of the theory in the previous section),
by replacing the leftmost $2N_{c}'$ of the $2N_{f}$ D6-branes in Figure 2
with an NS$'$5-brane. Then we take $N_{f}-N_{c}' \rightarrow N_{f}$
just as in the previous section, so that there should be $2N_{f}$
D6-branes, namely we should have $SU(2N_{f})$ global symmetry. We can identify
the fields (\ref{mfield}) and the tree-level superpotential (\ref{msuperpot})
in the magnetic theory from the resulting brane configuration
depicted in Figure 4.
\begin{figure}
    \centerline{\psfig{figure=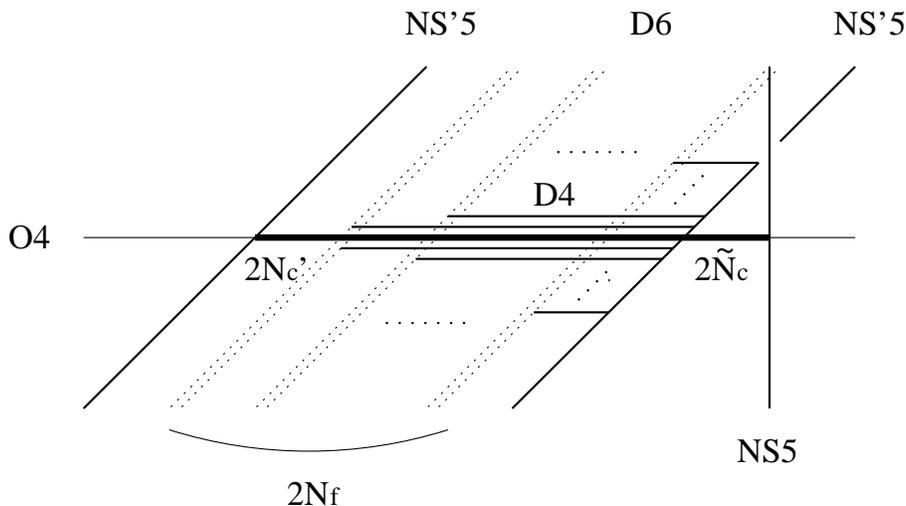,width=12cm}}
    \caption{Type IIA brane configuration for the magnetic
             $Sp(\tilde{N}_{c}) \times SO(2N_{c}')$ SQCD.}
\end{figure}

In the brane language, the duality transformation to the electric theory
with respect to $Sp(\tilde{N}_{c})$ gauge group corresponds to
translating the NS5-brane into between the left NS$'$5-brane and
the leftmost D6-brane. Performing the procedure explained in
section 2 in reverse order, we have the type IIA brane configuration for
the electric theory as shown in Figure 5.
\begin{figure}
    \centerline{\psfig{figure=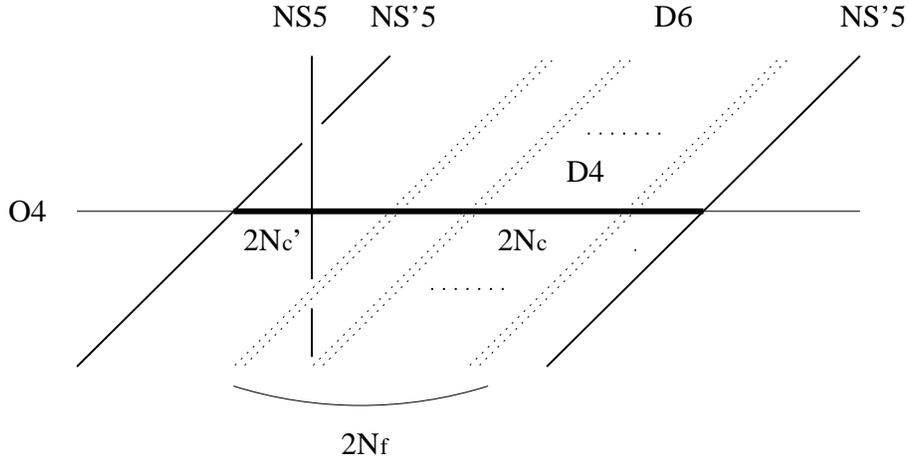,width=12cm}}
    \caption{Type IIA brane configuration for the electric
             $Sp(N_{c}) \times SO(2N_{c}')$ SQCD.}
\end{figure}

Let us now construct the corresponding M5-brane
configuration by rotating the $N=2$ configuration which is given by
replacing both of the two NS$'$5-branes with two NS5-branes. 
For a finite mass $\mu_a$ for the adjoint chiral superfields in
$N=2$ $Sp(N_c)$ and $SO(2N_c')$ vector multiplets,
the M5-brane's boundary conditions are given by
\begin{equation}
\label{boundary}
\begin{array}{rl}
{\rm (I)}   & v \rightarrow \infty, ~w \sim \mu_a v,
              ~y \sim \Lambda_{N=2}^{\prime -4(N_c'-1)+2N_c}v^{2(N_c'-1)}, \\
{\rm (II)}  & v \rightarrow \infty,~w \sim 0,
              ~y \sim v^{2N_c+2-2N_c'}, \\
{\rm (III)} & v \rightarrow \infty, ~w \sim \mu_a v,
              ~x \sim \Lambda_{N=2}^{-4(N_c+1)+2(N_f+N_c')}v^{2(N_c+1)}, \\
{\rm (IV)}  & y \rightarrow \infty, ~w^2 \sim 0, ~v^2 \sim 0,
\end{array}
\end{equation}
where $\Lambda_{N=2}$ and $\Lambda'_{N=2}$ are the dynamical scales for
$Sp(N_c)$ and $SO(2N_c')$ gauge groups in the presence of
$N=2$ supersymmetry. From the type IIA point of view, the conditions
(I), (II) and (III) characterize the bending pattern of the three NS5-branes
due to D4-branes' tension, and the condition (IV) indicates that
the orbifold fixed plane is screened by an M5-brane and its mirror image
in the region $y \sim \infty$, ensuring that the O4-plane has RR charge $+1$
on the left of the leftmost NS5-brane. Generally, the rotated curve is
parametrized by two holomorphic equations: $F(y,v)=0$ which characterizes
the $N=2$ configuration before the rotation, and $w=G(y,v)$ which
implies that $w$ is a meromorphic function over the Riemann surface.
If the M5-brane which has the asymptotic behaviour (\ref{boundary})
consists of several components, we must determine the two functions $f$
and $g$ for each of the components. We consider the following five
classes distinguished by the number of M5-brane components which satisfy the
first three boundary conditions, (I), (II) and (III). In the analysis,
we regard the curves as compactified Rieman surfaces by adding a point to
each of the asymptotic regions (I), (II) and (III).

\bigskip
{\it one component case}

Let us first examine if it is possible to enable only one component of
M5-brane to fulfill the three conditions (I)$\sim$(III) simultaneously.
We introduce the complex parameter $z=w-\mu_a v$ which is, similarly to
$w$, a meromorphic function over the surface. Then, the conditions
(I)$\sim$(III) inform us that $z$ has a simple pole at (II) and is
finite at all the other points. The existence of such a meromorphic
function forces the surface to be ${\bf P}^1$ and $z$ to be a global coordinate
parametrizing the entire surface, ${\bf P}^1$. However, this is
incompatible with the boundary conditions and ${\bf Z}_2$ symmetry
which require $z$ to have the same value $z=0$ at the two distinct
points (I) and (III) on the surface. Therefore, we conclude that there
is no holomorphically embedded single M5-brane with the boundary
behaviour (I)$\sim$(III).

\bigskip
{\it two components case 1}

We consider here the case in which there are two components,
$C_{{\rm I}}$ satisfying (I) and $C_{{\rm II-III}}$ satisfying (II) and
(III) simultaneously
\footnote{
In the following of this paper, the $C$'s with roman subscripts
refer to the M5-brane components which satisfy the boundary
conditions specified by the subscripts.
}.
Let us begin with $C_{{\rm I}}$. $C_{{\rm I}}$ is
parametrized by two equations of the forms $y=f(v), w=g(v)$, where $f$ and $g$
are some meromorphic functions. If $f$ has a zero at a finite value of 
$v = v_0 \neq 0$, $x$ goes to infinity for $v \rightarrow v_0$ as can be 
seen from (\ref{Taub-NUT}). This means that a semi-infinite D4-brane is
stuck on the rightmost NS5-brane from the right, which is not
contained in the type IIA brane configuration we started with. Thus, $f$ 
can have zeroes only at $v = 0$ or $v = \infty$. Besides, $f$ can have poles
only at $v = 0$ or $v = \infty$ so that there should be no D4-brane extending
semi-infinitely towards $y \rightarrow \infty$. $f$ is therefore a
(non-negative or negative) power of $v$. $g$ must also be a power of $v$ to avoid
unexpected boundary behaviour of the M5-brane. These restrictions
uniquely determine $C_{{\rm I}}$,
\begin{equation}
    C_{{\rm I}}\left\{
            \begin{array}{l}
              y=\Lambda_{N=2}^{\prime -4(N_c'-1)+2N_c}v^{2(N_c'-1)}, \\
              w=\mu_a v.
            \end{array} \right.
\end{equation}

We now turn to $C_{{\rm II-III}}$. The same reasoning as explained in the
previous case applied to the boundary conitions (II) and (III) shows
that $C_{{\rm II-III}}$ is a ${\bf P}^1$ with well-defined coordinate $w$.
Hence, $v$ and $x$ are expressed as meromorphic functons of $w$, which
are determined by the boundary conditions and the ${\bf Z}_2$ symmetry as
\begin{equation}
    C_{{\rm II-III}}\left\{
            \begin{array}{l}
               v=\mu_a^{-1}w^{-1}(w^2-w_0^2), \\
               x=\mu_a^{-2(N_c+1)}\Lambda_{N=2}^{-4(N_c+1)+2(N_f+N_c')}
                 w^{2(N_c+1)-2(N_f+N_c')}(w^2-w_0^2)^{N_f+N_c'},
            \end{array} \right.
\end{equation}
where $(-w_0^2)^{N_f+N_c'-2(N_c+1)}=\mu_a^{2(N_f+N_c')-4(N_c+1)}
\Lambda_{N=2}^{-4(N_c+1)+2(N_f+N_c')}$.

Since both of the components $C_{{\rm I}}$ and $C_{{\rm II-III}}$ do not 
satisfy the boundary condition (IV), we require an additional component
$C_{{\rm IV}}$,
\begin{equation}
    C_{{\rm IV}}\left\{
            \begin{array}{l}
               x_1^2=0, \\
               w=0.
            \end{array} \right.
\end{equation}
Although $C_{{\rm I}}$ intersects with the infinite orbifold fixed plane
$C_y$ transversely for $N_c'=1$, supersymmetry is not broken because
the component $C_{{\rm IV}}$ is a pair of M5-brane screening $C_y$.
Since all the three components $C_{{\rm I}}$, $C_{{\rm II-III}}$ and $C_{{\rm
IV}}$ are holomorphic, they define supersymmetric vacua for
a {\it finite} $\mu_a$.

In order to obtain the M5-brane configuration corresponding to the electric
theory with the adjoint fields completely decoupled, we take $\mu_a
\rightarrow \infty$ limit fixing the $N=1$ scales $\Lambda$ and
$\Lambda'$ which are related to the $N=2$ scales by
\begin{equation}
  \label{LambdaN=2}
  \Lambda^{3(N_c+1)-(N_f+N_c')}
    =\mu_a^{N_c+1}\Lambda_{N=2}^{2(N_c+1)-(N_f+N_c')},
\end{equation}
\begin{equation}
  \label{Lambda'N=2}
  \Lambda^{\prime 6(N_c'-1)-2N_c}
    =\mu_a^{2(N_c'-1)}\Lambda_{N=2}^{\prime 4(N_c'-1)-2N_c}.
\end{equation}
For $N_c=-1$ which realizes the IYIT model with gauged global symmetry,
$(-w_0^2)^{N_f+N_c'}=(\mu_a^2 \Lambda^2)^{N_f+N_c'}$ diverges as
$\mu_a \rightarrow \infty$. Therefore, the curve $C_{{\rm II-III}}$ is
infinitely elongated, implying that the supersymmetric vacua which exist
for a finite adjoint mass $\mu_a$ run away towards far infinity in the
limit $\mu_a \rightarrow \infty$.

\bigskip
{\it two components case 2}

Let us consider the M5-brane configuration consisting of two components
$C_{{\rm II}}$ and $C_{{\rm I-III}}$. Following the argument on $C_{{\rm
I}}$ given in the previous case, $C_{{\rm II}}$ can be determined as
\begin{equation}
  \label{CII}
    C_{{\rm II}}\left\{
              \begin{array}{l}
               y=v^{2N_c+2-2N_c'}, \\
               w=0.
              \end{array} \right.
\end{equation}
The holomorphic function $F(y,v)$ for $C_{{\rm I-III}}$ is a second
order polynomial of $y$ with its two roots behave like (I) and (III) for
$v \rightarrow \infty$. We thus obtain
\begin{equation}
  \label{CI-III}
   C_{{\rm I-III}}\left\{
              \begin{array}{l}
               \Lambda_{N=2}^{\prime 4(N_c'-1)-2N_c}y^2
               -P_{N_c'-1}(v^2)y
               +\Lambda_{N=2}^{4(N_c+1)-2(N_f+N_c')}v^{2(N_f+N_c'-N_c-2)}=0, \\
                w=\mu_a v,
              \end{array} \right.
\end{equation}
where $P_{N_c'-1}(v^2)=v^{2(N_c'-1)}+\sum_{n=0}^{N_c'-2}a_nv^{2n}$ is a
polynomial of the $(N_c'-1)$-th order.

Let us put $N_c=-1$ in (\ref{CII}). Then, $C_{{\rm II}}$ satisfies the
condition (IV) beside (II) for $N_c'=1$, while $C_{{\rm II}}$ screens
$C_y$ by more than two M5-branes for $N_c' \geq 2$, being imcompatible
with the condition (IV). Putting $N_c=-1,N_c'=1$ and taking the limit
$\mu_a \rightarrow \infty$ with the $N=1$ scales given in
(\ref{LambdaN=2}) and (\ref{Lambda'N=2}) fixed, we find that
$C_{{\rm I-III}}$ breaks up into the two components $C_{{\rm I}}$ and
$C_{{\rm III}}$. 

The resultant configuration for $N_c'=1$ is thus
\begin{equation}
  \label{CI-CII-CIII}
  C_{{\rm I}}\left\{
            \begin{array}{l}
              y=\Lambda^{\prime -2}, \\
              v=0,
            \end{array} \right. \hspace{1cm}
  C_{{\rm II}}\left\{
            \begin{array}{l}
              y=v^{-2}, \\
              w=0,
            \end{array} \right. \hspace{1cm}
  C_{{\rm III}}\left\{
            \begin{array}{l}
              x=\Lambda^{2(N_f+1)}, \\
              v=0,
            \end{array} \right.
\end{equation}
and there is no holomorphic configuration for $N_c' \geq 2$.

\bigskip
{\it two components case 3}

The last possibility with two M5-brane components is that
$C_{{\rm III}}$ and $C_{{\rm I-II}}$ constitute the entire
configuration. We can determine the forms of $C_{{\rm III}}$ and
$C_{{\rm I-II}}$ in the same manner as in the previous cases,
\begin{equation}
  \label{CIII}
    C_{{\rm III}}\left\{
            \begin{array}{l}
              x=\Lambda_{N=2}^{-4(N_c+1)+2(N_f+N_c')}v^{2(N_c+1)}, \\
              w=\mu_a v,
            \end{array} \right.
\end{equation}
\begin{equation}
  \label{CI-II}
    C_{{\rm I-II}}\left\{ \begin{array}{l}
               v=\mu_a^{-1}w^{-1}(w^2-w_0^{\prime 2}), \\
               y=\mu_a^{-2(N_c'-1)}\Lambda_{N=2}^{\prime -4(N_c'-1)+2N_c}
                 w^{2(N_c'-1)-2N_c}(w^2-w_0^{\prime 2})^{N_c},
            \end{array} \right.
\end{equation}
where $(-w_0^{\prime 2})^{N_c-2(N_c'-1)}=\mu_a^{2N_c-4(N_c'-1)}
\Lambda_{N=2}^{\prime -4(N_c'-1)+2N_c}$. However, putting $N_c=-1$ in
(\ref{CI-II}) indicates that $y \rightarrow \infty$ for $w^2 \rightarrow
w_0^{\prime 2}$. Therefore, this possibility is excluded by the
requirement that no semi-infinite D4-brane is present.

\bigskip
{\it three components case}

It is also possible that the M5-brane configuration consists of three
components $C_{{\rm I}}$, $C_{{\rm II}}$ and $C_{{\rm III}}$. As far as
$N_c=-1$, this configuration is consistent with the boundary condition
(IV) only for $N_c'=1$. For $N_c=-1,N_c'=1$, the three components reduce 
to (\ref{CI-CII-CIII}) in the limit $\mu_a \rightarrow \infty$.

\bigskip
To summarize, there exists no holomorphic M5-brane configuration for the
IYIT model with gauged $SO(2N_c')$ ($N_c' \geq 2$) subgroup of the
global symmetry. On the other hand, if the smallest subgroup $U(1) \simeq
SO(2)$ is gauged, we can construct uniquely the holomorphic configuration
(\ref{CI-CII-CIII}) which we depict in Figure 6.
\begin{figure}
    \centerline{\psfig{figure=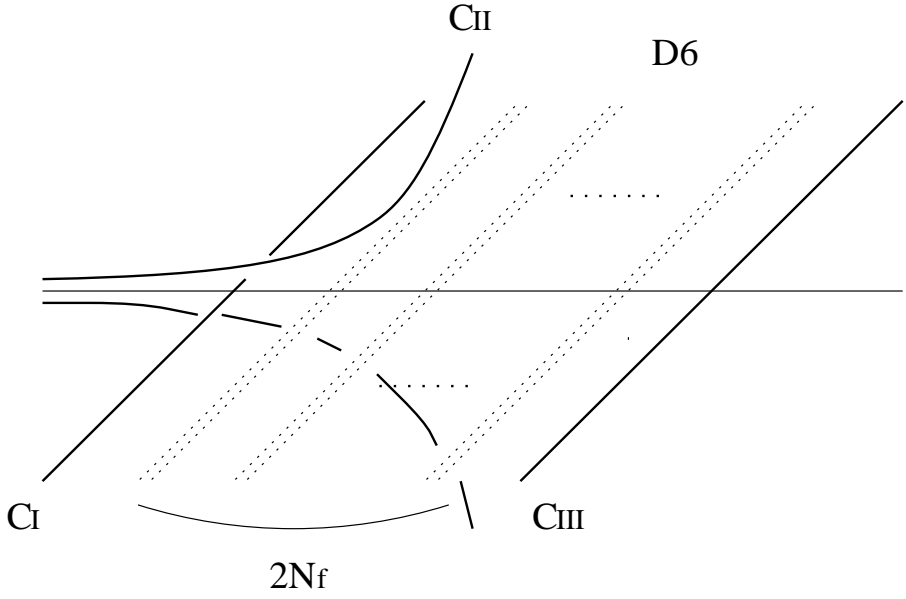,width=12cm}}
    \caption{M5-brane configuration for the electric
             $Sp(-1) \times U(1)$ gauge theory.}
\end{figure}
Since $C_{{\rm I}}$ and $C_{{\rm III}}$ is intersecting transversely
with $C_y$ and $C_x$ respectively, the theory posseses a stable vacuum with 
dynamically broken supersymmetry.

Finally, we check the validity of constructing M5-brane configuration from the 
electric $Sp(-1)$ theory, by demonstrating that the configuration
(\ref{CI-CII-CIII}) can be smoothly deformed to the one which
resembles the magnetic type IIA configuration shown in Figure 4.
We expect that the magnetic description becomes manifest for a small
$Sp(\tilde{N_c})$ scale $\tilde{\Lambda}$.
For the purpose of taking the limit $\tilde{\Lambda} \rightarrow 0$,
it is convenient to reparametrize the variables $x$ and $y$ in the way 
\begin{equation}
  \label{newparam}
  \left\{
        \begin{array}{l}
          \tilde{x}=\Lambda^{6(N_c+1)-2(N_f+N_c')}x, \\
          \tilde{y}=\Lambda^{-6(N_c+1)+2(N_f+N_c')}y.
        \end{array} \right.
\end{equation}
Under the condition  $N_c=-1$ and $N_c'=1$ we are now interested in,
$C_{{\rm I}}$, $C_{{\rm II}}$ and $C_{{\rm III}}$ in (\ref{CI-CII-CIII})
are expressed in terms of the new parameters $\tilde{x},\tilde{y}$ as
\begin{equation}
  C_{{\rm I}}\left\{
        \begin{array}{l}
          \tilde{y}=(\mu^{-1}\tilde{\Lambda}^{\prime 2})^{2\tilde{N}_c}, \\
          v=0,
        \end{array} \right. \hspace{1cm}
  C_{{\rm II}}\left\{
        \begin{array}{l}
          \tilde{y}=(\mu^{-1}\tilde{\Lambda}^2)^{2(\tilde{N}_c+1)}v^{-2}, \\
          w=0,
        \end{array} \right. \hspace{1cm}
  C_{{\rm III}}\left\{
        \begin{array}{l}
          \tilde{x}=1, \\
          v=0,
        \end{array} \right.
\end{equation}
where we have used the relation between the electric and the magnetic
scales, (\ref{Lambda}) and (\ref{Lambda'}). We take the magnetic limit by
pushing $\tilde{\Lambda} \rightarrow 0$ while the magnetic $SO(2N_c')$
scale $\tilde{\Lambda}'$ is fixed, that is, we deform only the component
$C_{{\rm II}}$. The first equation in $C_{{\rm II}}$ is written in terms 
of the coordinates $(\tilde{x}_i,\tilde{y}_i)$
\footnote{
The coordinates $(\tilde{x}_i,\tilde{y}_i)$ are related to the original
ones $(x_i,y_i)$ in the same way as (\ref{newparam}).
}
of the patch $U_i$ in the Taub-NUT space as
\begin{equation}
  \tilde{x}_i^{i+1}\tilde{y}_i^{i+2}
      =(\mu^{-1}\tilde{\Lambda}^2)^{2(\tilde{N}_c+1)}.
\end{equation}
From this, we learn that $C_{{\rm II}}$ is approximately wrapped on
$C_y,C_1,C_2,\cdots,C_{2N_f-1},C_x$ with multiplicities
$2,3,4,\cdots,2N_f+1,2N_f+2$, for a sufficiently small
$\tilde{\Lambda}$. As shown in Figure 7, the M5-brane configuration in
the magnetic limit correctly reproduces the magnetic type IIA
configuration with $N_f+N_c'=\tilde{N}_c+1$ and $N_c'=1$.

\begin{figure}
    \centerline{\psfig{figure=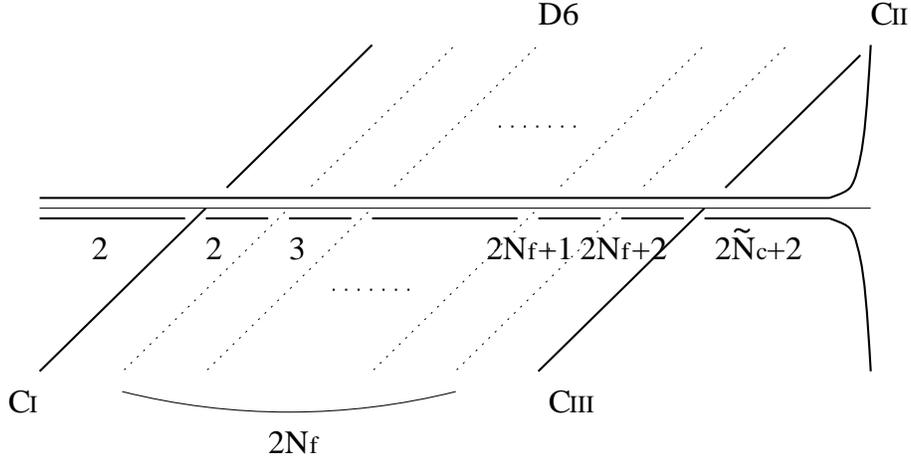,width=12cm}}
    \caption{M5-brane configuration for the magnetic
             $Sp(\tilde{N}_c) \times U(1)$ gauge theory.}
\end{figure}

\section{Conclusions}
\setcounter{equation}{0}

We have investigated $Sp(N_c) \times SO(2N_c')$ SQCD which is obtained
by gauging $SO(2N_c')$ subgroup of the flavour symmetry in the IYIT
model. For non-Abelian gauge group $SO(2N_c')$ $(N_c' \geq 2)$, the
M5-brane configuration is necsssarily non-holomorphic. Nevertheless, it
is difficult in the brane framework to verify the presence of
supersymmetry breaking local minima as stressed in section 2. On the
contrary, in the case of gauged $U(1)$ symmetry we have reached the
conclusion that the supersymmetry breaking stable vacuum exists, because
the corresponding M5-brane configuration cannot avoid inconsistent
configurations which we have encountered also in the original IYIT
model. This observation will support the idea to construct the
gauge-mediated models based on the IYIT model which plays the role of
the hidden sector.

So far, the analyses of supersymmetric gauge theories via branes in
string theories or M-theory have contributed to the understanding of
{\it brane dynamics} for the most part, not of {\it field theories}
themselves. That is, while we have successfully derived various
properties of branes such as s-rule \cite{HM,OV}, the inconsistency of
t-configuraton, etc. to account for the field theory results, there are
few known examples of new consequences on field theories which are
extracted from brane configurations. As a counterexample to this, we have
proposed in this paper that branes can provide us with a new method for
searching the models exhibiting DSB. However, t-configuration to which
we have attributed DSB in field theories, is presumably one example of
the inconsistent brane configurations which cause supersymmetry to be broken
spontaneously. Studying the other models which are known to realize DSB
in field theories, we may find other kinds of inconsistent configuration which 
allow us to discover new DSB models as we have achieved in this
work.

\section*{Acknowledgment}

We would like to thank K.~Hori, Izawa K.-I., Y.~Nomura and T.~Yanagida
for very useful comments and careful reading of the manuscript.

\newpage
%
%
%
\newcommand{\Journal}[4]{{#1} {\bf #2} {(#3)} {#4}}
\newcommand{\APJ}{Ap. J.} \newcommand{\CJP}{Can. J. Phys.}
\newcommand{\NC}{Nuovo Cimento} \newcommand{\NP}{Nucl. Phys.}
\newcommand{\PL}{Phys. Lett.} \newcommand{\PR}{Phys. Rev.}
\newcommand{\PRep}{Phys. Rep.} \newcommand{\PRL}{Phys. Rev. Lett.}
\newcommand{\PTP}{Prog. Theor. Phys.} \newcommand{\SJNP}{Sov. J. Nucl.
  Phys.}
\newcommand{\ZP}{Z. Phys.}

%
%


\begin{thebibliography}{99}
%
      \bibitem{PT} For a review, see
        E.~Poppitz and S.~Trivedi,
        hep-th/9803107.
%
      \bibitem{ADS}
        I.~Affleck, M.~Dine and N.~Seiberg,
        \Journal{\NP}{B241}{1984}{493},
        \Journal{\NP}{B256}{1985}{557}.
%
      \bibitem{IYIT}
        Izawa~K.-I. and T.~Yanagida,
        \Journal{\PTP}{95}{1996}{829};\\
        K.~Intriligator and S.~Thomas,
        \Journal{\NP}{B473}{1996}{121}.
%
      \bibitem{ISS}
        K.~Intriligator, N.~Seiberg and S.~H.~Shenker,
        \Journal{\PL}{B342}{1995}{152}.
%
      \bibitem{IS1}
        N.~Seiberg,
        \Journal{\PR}{D49}{1994}{6857},
        \Journal{\NP}{B435}{1995}{129}; \\
        K.~Intriligator and N.~Seiberg,
        \Journal{Nucl. Phys. Proc. Suppl.}{45BC}{1996}{1}.
%
      \bibitem{IS2}
        K.~Intriligator and N.~Seiberg,
        \Journal{\NP}{B444}{1995}{125}.
%
      \bibitem{IP}
        K.~Intriligator and P.~Pouliot,
        \Journal{\PL}{B353}{1995}{471}.
%
      \bibitem{GK} For a conprehensive review, see
        A.~Giveon and D.~Kutasov,
        hep-th/9802067.
%
      \bibitem{Tseytlin}
        A.~A.~Tseytlin,
        \Journal{\NP}{B469}{1996}{51}.
%
      \bibitem{GG}
        M.~B.~Green and M.~Gutperle,
        \Journal{\PL}{B377}{1996}{28}.
%
      \bibitem{Witten1}
        E.~Witten,
        \Journal{\NP}{B500}{1997}{3}.
%
      \bibitem{BSTY1}
        A.~Brandhuber, J.~Sonnenschein, S.~Theisen and S.~Yankielowicz,
        \Journal{\NP}{B502}{1997}{125}.
%
      \bibitem{BSTY2}
        A.~Brandhuber, J.~Sonnenschein, S.~Theisen and S.~Yankielowicz,
        \Journal{\NP}{B504}{1997}{175}.
%
      \bibitem{LLL1}
        K.~Landsteiner, E.~Lopez and D.~A. Lowe,
        \Journal{\NP}{B507}{1997}{197}.
%
      \bibitem{AOT1}
        C.~Ahn, K.~Oh and R.~Tatar,
        hep-th/9708127.
%
      \bibitem{Terashima}
        S.~Terashima,
        \Journal{\NP}{B526}{1998}{163}.
%
      \bibitem{LLL2}
        K.~Landsteiner, E.~Lopez and D.~A. Lowe,
        \Journal{J. High Energy Phys.}{07}{1998}{011}.
%
      \bibitem{Witten2}
        E.~Witten,
        \Journal{Nucl. Phys. Proc. Suppl.}{68}{1998}{216}.
%
      \bibitem{HOO}
        K.~Hori, H.~Ooguri, and Y.~Oz,
        \Journal{Adv. Theor. Math. Phys.}{1}{1998}{1}.
%
      \bibitem{BHOO}
        J.~de~Boer, K.~Hori, H.~Ooguri and Y.~Oz,
        \Journal{\NP}{B522}{1998}{20}.
%
      \bibitem{AOT2}
        C.~Ahn, K.~Oh and R.~Tatar,
        hep-th/9803197.
%
      \bibitem{EGK}
        S.~Elitzur, A.~Giveon and D.~Kutasov,
        \Journal{\PL}{B400}{1997}{269}.
%
      \bibitem{EJS}
        N.~Evans, C.~V.~Johnson and A.~D.~Shapere,
        \Journal{\NP}{B505}{1997}{251}.
%
      \bibitem{BH}
        J.~H.~Brodie and A.~Hanany,
        \Journal{\NP}{B506}{1997}{157}.
%
      \bibitem{EGKRS}
        S.~Elitzur, A.~Giveon, D.~Kutasov, E.~Rabinovici and A. Schwimmer,
        \Journal{\NP}{B505}{1997}{202}.
%
      \bibitem{CS}
        C.~Csaki and W.~Skiba,
        \Journal{\PL}{B415}{1997}{31}.
%
      \bibitem{Sugimoto}
        S.~Sugimoto,
        hep-th/9804114.
%
      \bibitem{Hori1}
        K.~Hori,
        hep-th/9805142.
%
      \bibitem{Hori2}
        K.~Hori,
        hep-th/9805141.
%
      \bibitem{GR}
        For a review, G.~F.~Giudice and R.~Rattazzi,
        hep-ph/9801271.
%
      \bibitem{gm}
        M.~Dine, A.~E.~Nelson and Y.~Shirman,
        \Journal{\PR}{D51}{1995}{1362}; \\
        M.~Dine, A.~E.~Nelson, Y.~Nir and Y.~Shirman,
        \Journal{\PR}{D53}{1996}{2658}; \\
        M.~Dine, Y.~Nir and Y.~Shirman,
        \Journal{\PR}{D55}{1997}{1501}.
%
      \bibitem{INTY}
        Izawa~K.-I., Y.~Nomura, K.~Tobe and T.~Yanagida,
        \Journal{\PR}{D56}{1997}{2886}; \\
        Y.~Nomura and K.~Tobe,
        \Journal{\PR}{D58}{1998}{055002}; \\
        Y.~Nomura, K.~Tobe and T.~Yanagida,
        \Journal{\PL}{B425}{1998}{107}.
%
      \bibitem{EGKT}
        S.~Elitzur, A.~Giveon, D.~Kutasov and D.~Tsabar,
        \Journal{\NP}{B524}{1998}{251}.
%
      \bibitem{PST}
        E.~Poppitz, Y.~Shadmi and S.~P.~Trivedi,
        \Journal{\NP}{B480}{1996}{125},
        \Journal{\PL}{B388}{1996}{561}.
%
      \bibitem{AOT3}
        C.~Ahn, K.~Oh and R.~Tatar,
        hep-th/9707027.
%
      \bibitem{LO}
        E. Lopez and B. Ormsby,
        hep-th/9808125.
%
      \bibitem{AM}
        N.~Arkani-Hamed and H.~Murayama,
        \Journal{\PR}{D57}{1998}{6638}.
%
      \bibitem{HM}
        A.~Hanany and E.~Witten,
        \Journal{\NP}{B492}{1997}{152}.
%
      \bibitem{OV}
        H.~Ooguri and C.~Vafa,
        \Journal{\NP}{B500}{1997}{62}.
%
\end{thebibliography}
\end{document}